\documentclass{article}

\usepackage{times}
\usepackage{graphicx} % more modern
\usepackage{subfigure} 

\usepackage{natbib}

\usepackage{algorithm}
\usepackage{algorithmic}
\usepackage{booktabs,hhline,tabularx,multirow}

\usepackage[accepted,nohyperref]{icml2017}

\icmltitlerunning{Multi-Level and Multi-Scale Feature Aggregation Using Pre-trained Sample-level Deep Convolutional Neural Networks}

\begin{document} 

\twocolumn[
\icmltitle{Multi-Level and Multi-Scale Feature Aggregation Using Sample-level Deep Convolutional Neural Networks for Music Classification}



% you can specify symbols, otherwise they are numbered in order

\begin{icmlauthorlist}
\icmlauthor{Jongpil Lee}{kaist}
\icmlauthor{Juhan Nam}{kaist}
\end{icmlauthorlist}

\icmlaffiliation{kaist}{Graduate School of Culture Technology, Korea Advanced Institute of Science and Technology (KAIST), Daejeon, Korea}

\icmlcorrespondingauthor{Juhan Nam}{juhannam@kaist.ac.kr}

\icmlkeywords{convolutional neural networks, feature aggregation, music auto-tagging, transfer learning, raw waveforms, sample-level filters}

\vskip 0.3in
]



\printAffiliationsAndNotice{} % otherwise use the standard text.

\begin{abstract} 
%Individual music description words have different local and global characteristics of music audio.
Music tag words that describe music audio by text have different levels of abstraction. 
%To take this problem into account, we combine two methods of dealing with this problem at different levels, which are multi-level and multi-scale feature aggregation and sample-level deep convolutional neural networks. 
Taking this issue into account, we propose a music classification approach that aggregates multi-level and multi-scale features using pre-trained feature extractors. In particular, the feature extractors are trained in sample-level deep convolutional neural networks using raw waveforms. We show that this approach achieves state-of-the-art results on several music classification datasets.
\end{abstract}

\section{Introduction}
\label{introduction}

%Content based music representation learning is very important to overcome popularity bias problem, cold start problem, and dependency on listening data which are disadvantages of user data based recommendation system. 

Learning hierarchical audio representations for music classification in an end-to-end manner is a challenge due to the diversity of music description words. In this study, we combine two previously proposed methods to tackle the problem.   

% due to the diversity of music description words and well-established frame-level time-frequency representations
 
\subsection{Multi-Level and Multi-Scale Feature Aggregation}
Music classification tasks, particularly music auto-tagging among others, have a wide variety of labels in terms of genre, mood, instruments and other song characteristics. In order to address different levels of abstraction that the labels retain, we recently proposed an approach that aggregates audio features extracted in a multi-level and multi-scale manner \cite{lee2017multi}. 
% Since individual music descriptions have different audio characteristic sensitivity to different time scales and levels of features, in Lee \& Nam's work \yrcite{lee2017multi}, they proposed a CNN-based method that handles multi-level and multi-scale of audio features. 
The method is composed of three steps: extracting features using pre-trained convolutional neural networks (CNNs), feature aggregation and song-level classification. The CNNs are trained in a supervised manner with the tag labels, taking different sizes of input frames. The feature aggregation step extracts multiple-level features using the pre-trained CNNs and  summarizes them into a single song-level feature vector. The last step performs final predictions of tags from the aggregated features using a fully-connected neural network. 
This multi-step architecture has the advantage of capturing local and global characteristics of a song and also has a good accordance with transfer learning. However, our previous approach used mel-frequency spectrograms as input, which are based on the knowledge of pitch perception.

% have limitations in using fixed window and hop size mel-spectrogram CNN models that cannot fit into the low-level characteristics of individual tag.
% by applying frame-wise max-pooling and segment-wise average-pooling on the intermediate convolutional layers. The song-level classification makes final prediction using the aggregated features. This method has the advantage of capturing local and global characteristics of the entire song and summarizing music audio of different lengths.

\begin{figure}[t]
\vskip -0.2in
\begin{center}
\centerline{\includegraphics[width=\columnwidth]{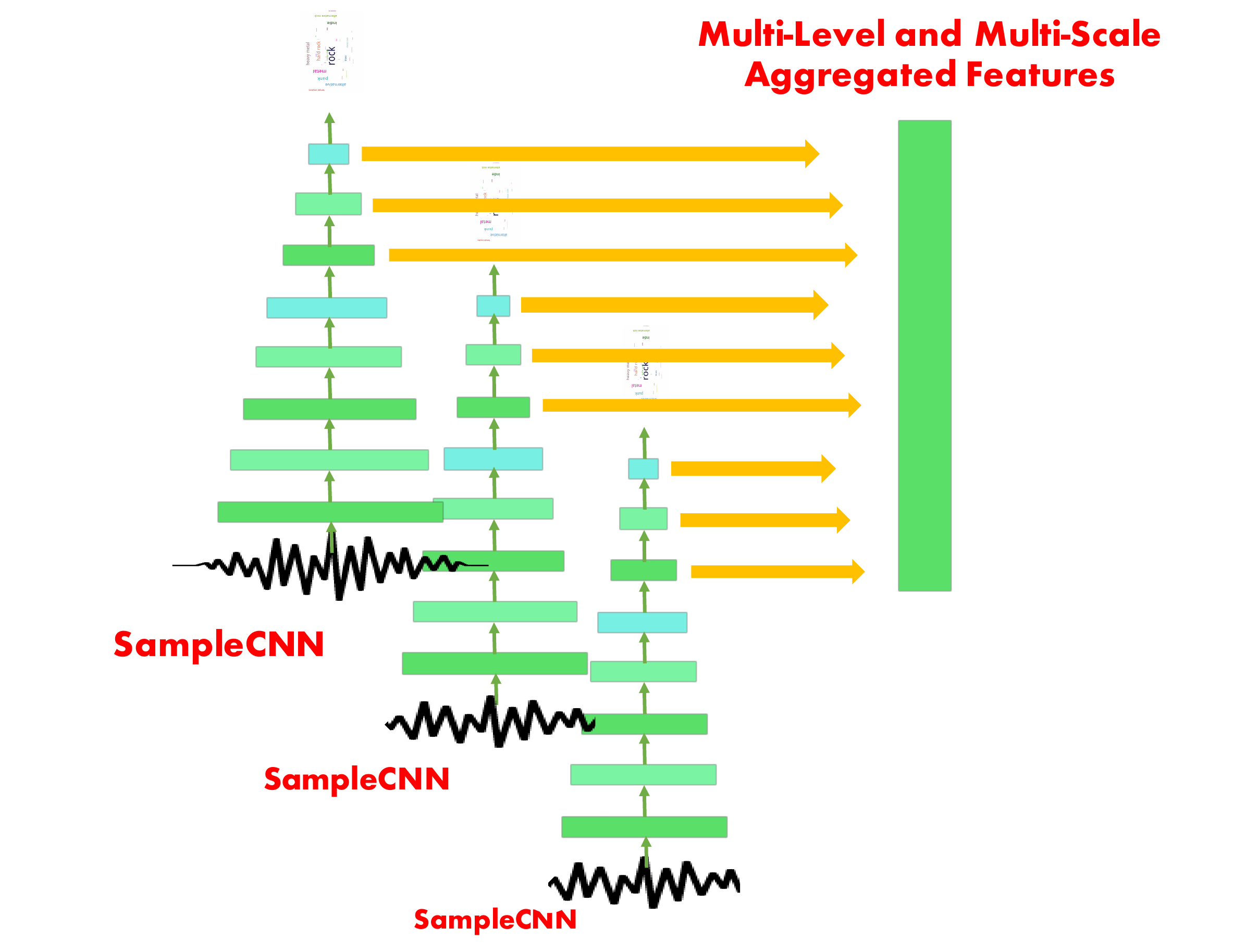}}
\caption{Multi-level and multi-scale feature aggregation using sample-level deep convolutional neural networks}
\label{figure1}
\end{center}
\vskip -0.2in
\end{figure}

\begin{table*}[ht]
\vspace{-3.3mm}
\caption{Comparison with previous work. Note that we used only multi-level features in the proposed work due to the long training time with MSD (it took about two weeks on the GTX1080Ti GPU to train the DCNN and so we used only the $ 3 ^ 9 $ model). Also, "-$n$ LAYER" indicates $n$ layers below from the output.}
\label{table1}
\vskip 0.15in
\begin{center}
\begin{small}
\begin{sc}
\begin{tabular}{cccccc}
\multicolumn{2}{c}{MODEL} & \begin{tabular}[c]{@{}c@{}} GTZAN\\(Acc.) \end{tabular} & \begin{tabular}[c]{@{}c@{}} MTAT\\(AUC) \end{tabular} & \begin{tabular}[c]{@{}c@{}} TAGTRAUM\\(Acc.) \end{tabular} & \begin{tabular}[c]{@{}c@{}} MSD\\(AUC) \end{tabular} \\
\hline
\abovespace\belowspace
\cite{lee2017multi} &  \begin{tabular}[c]{@{}c@{}} multi-level and multi-scale features \\ (pre-trained with MSD) \end{tabular}  & 0.720 & 0.9021  &  0.766  &  \textbf{0.8878} \\
\belowspace
\cite{lee2017sample}&Sample-level DCNN ($3^9$ model) & - & 0.9055 & - & 0.8812 \\
\hline
\abovespace\belowspace
Proposed Work & -3 layer (pre-trained with MSD)   & 0.778& 0.8988& 0.760& 0.8831 \\
\belowspace
(features from& -2 layer (pre-trained with MSD) & 0.811& 0.8998& 0.768& 0.8838\\
\belowspace
sample-level& -1 layer (pre-trained with MSD)    & \textbf{0.821}& 0.8976& 0.768& 0.8842 \\
\belowspace
dcnn, $3^9$ model)& last 3 layers (pre-trained with MSD)   & 0.805& 0.9018& \textbf{0.768}& 0.8842 \\
\hline
\end{tabular}
\end{sc}
\end{small}
\end{center}
\vskip -0.1in
\end{table*}

\begin{table}[ht]
%\vspace{1.6mm}
\caption{Comparison of various multi-scale feature combinations. Only MTAT was used.}
\label{table2}
\vskip 0.15in
\begin{center}
\begin{small}
\begin{sc}
\begin{tabular}{cc}
\begin{tabular}[c]{@{}c@{}} features from sample-level DCNNs\\ Last 3 layers \\ (pre-trained with MTAT) \end{tabular} & MTAT \\
\hline
\abovespace\belowspace
$3^9$ model&   0.9046 \\
\belowspace
$3^8$ and $3^9$ models &  0.9061 \\
\belowspace
$2^{13}$, $2^{14}$, $3^8$ and $3^9$ models &   0.9061   \\
\belowspace
$2^{13}$, $2^{14}$, $3^8$, $3^9$, $4^6$, $4^7$, $5^5$ and $5^6$ models &  \textbf{0.9064}  \\
\hline
\end{tabular}
\end{sc}
\end{small}
\end{center}
\vskip -0.1in
\end{table}

\begin{figure}[ht]
%\vspace{-3.0mm}
%\vskip 0.2in
\begin{center}
\centerline{\includegraphics[width=\columnwidth]{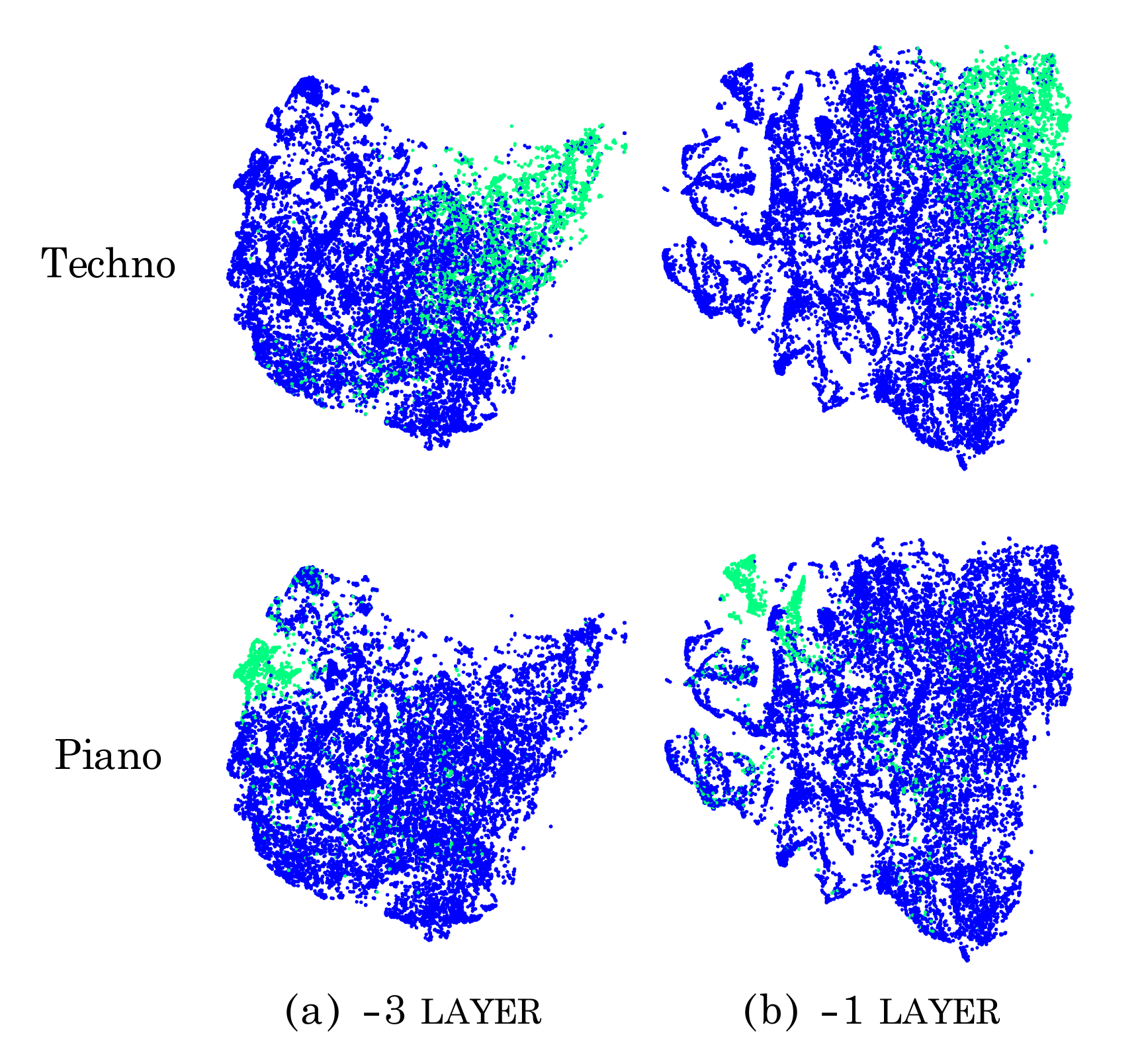}}
\caption{This figure shows feature visualization on songs with Piano tag and songs with Techno tag on MTAT using t-SNE. Features are extracted from (a) -3 LAYER and (b) -1 LAYER of the $3^9$ model pre-trained with MSD.} 
\label{figure2}
\end{center}
\vskip -0.3in
\end{figure} 

\subsection{Sample-level Deep Convolutional Neural Networks}
We recently investigated the possibility of employing raw waveforms as input for deep convolutional neural networks (DCNNs)  in music auto-tagging \cite{lee2017sample}. They were configured to take a very small grain of waveforms, even 2 or 3 samples, in the bottom-level filters. They show that the ``sample-level'' representation learning works well and the learned filters in each layer are sensitive to log-scaled frequency along layer such as mel-frequency spectrogram. 
% In mel-spectrogram based music classification model, considering diverse cases of mid-representation parameters (in audio pre-processing) is not trivial. Optimal parameters of spectrograms, nonlinear logarithmic scales (such as Mel) and amplitude compression can vary for individual tags. In Lee et al.'s recent work \yrcite{lee2017sample}, they proposed sample-level DCNN which learn representations from very small grains of waveforms (e.g. 2 or 3 samples) and also verified that learned filters in each layer are sensitive to log-scaled frequency along layer, such as mel-frequency spectrogram. 
% We should note that this architecture showed the state-of-the-arts performance without the need for tuning audio processing parameters and input normalization. This can be interpreted that these parameters have been learned for individual tags. 

\subsection{The combination}
In this study, we combine the two methods to take all the advantages of them. As illustrated in Figure \ref{figure1}, we used the top three hidden layers from the sample-level DCNNs for multi-level feature extraction. The DCNNs take different sizes of input. 

%each the two methods outlined above as

\section{Datasets}

We validate the effectiveness of the proposed method on different sizes of datasets for genre classification and auto-tagging. The details are as follows\footnote{https://github.com/jongpillee/music\_dataset\_split}:
\begin{itemize}
\item GTZAN (fault-filtered version) \cite{tzanetakis2002musical,kereliuk2015deep}: 930 songs, genre classification (10 genres) 
\item MagnaTAgaTune (MTAT) \cite{law2009evaluation}: 21105 songs, auto-tagging (50 tags) 
\item Million Song Dataset with Tagtraum genre annotations (TAGTRAUM, stratified
split with 80\% training data of CD2C version) \cite{schreiber2015improving}: 189189 songs, genre classification (15 genres)
\item Million Song Dataset with Last.FM tag  annotations (MSD) \cite{bertin2011million}: 241889 songs, auto-tagging (50 tags)
\end{itemize}

\section{Results \& Conclusion}

We obtained the results from the average of 10 experiments. From Table \ref{table1}, although the proposed method failed to outperform the best of the previous works on MSD and MTAT, the multi-level and multi-scale aggregation generally improves the performance. The improvement is particularly dominant in GTZAN. From Table \ref{table2} where only MTAT is used, the proposed method is superior to the two previous works. Furthermore, we visualize the features at different levels for selected tags in the Figure \ref{figure2}. Songs with genre tag (\emph{Techno}) are more closely clustered in the higher layer (-1 layer). On the other hand, songs with instrument tag (\emph{Piano}) are more closely clustered in the lower layer (-3 layer). This may indicate  that the optimal layer of feature representations can be different depending on the type of labels. All of these results show that the proposed feature aggregation method is also effective with the sample-level DCNNs. 

%Second, by combining the two methods, we could achieve state-of-the-arts performances on almost every datasets tested. Third, performance is expected to improve when using multi-scale features because we used only a single model pre-trained with MSD to extract features from the experiments in the table \ref {table1}. 

\bibliography{example_paper}
\bibliographystyle{icml2017}

\end{document}